# On the Stability of a Full-Duplex Aloha Network

Andrea Munari, *Member, IEEE*, Francesco Rossetto, *Senior Member, IEEE*,
Petri Mähönen, *Senior Member, IEEE*, Marina Petrova, *Member, IEEE*

*Abstract*— This letter offers the first characterisation of the stability for full-duplex large size networks. Through stochastic geometry tools, key performance metrics like delay and maximum stable arrival rate are derived for the non saturated system and compared to a half-duplex counterpart, also accounting for imperfect self-interference cancellation. This analysis better identifies that the full-duplex advantage is prominent for sparse networks, whereas in dense topologies residual self-interference may hinder achievable gains.

## I. INTRODUCTION

THE steady quest for capacity has recently drawn the attention of the wireless research community to the in-band full-duplex (FD) paradigm. Advances in self-interference cancellation (IC) techniques have in fact rendered simultaneous radio transmission and reception over the same bandwidth viable, and close-to-twofold capacity improvements for isolated links have been demonstrated [1]. In turn, design efforts at the physical layer have been flanked by analytical studies that cast light on the true potential and performance drivers of FD in more articulated scenarios. In particular, stochastic geometry models have characterised the key tradeoff between increased spatial reuse and additional interference due to simultaneous bi-directional links in large Aloha-based networks. Derived results show how in such settings FD can indeed provide remarkable throughput gains – although far from the alluring 2x factor – when used in lightly loaded networks [2] or for short-range and short-duration data exchanges [3]. However, existing works focus on saturated conditions in which terminals always have data to transmit, leaving the performance of more realistic buffered systems an open problem. From this standpoint, bringing the queueing behaviour of nodes into the picture triggers new and non-trivial questions. On the one hand, the number of bi-directional connections may be intrinsically limited by the lack of concurrent reciprocal traffic among neighbouring nodes. On the other hand, the achievable stability region and the delay performance will be critical decision factors for the success of the FD technology in large networks. This letter bridges this gap by offering the first characterisation of buffered Aloha systems with FD-capable terminals. The stability region as well as the throughput and delay are derived analytically in comparison to the ones of a half-duplex (HD) counterpart, and the effect of imperfect IC is discussed. The framework is verified by means of simulations.

## II. SYSTEM MODEL

Within this letter, we focus on a bi-dimensional wireless system with full-duplex capable nodes. On the $\mathbb{R}^2$ plane, we denote the position of a point through a vector $\mathbf{u}$. In polar coordinates, we expressed $\mathbf{u} = ue^{j\varphi}$, where $u = \|\mathbf{u}\|$ is the Euclidean norm and $\varphi \in [0, 2\pi)$. At the beginning of the observation period (time $t = 0$) the network topology is drawn as per a homogeneous independent Poisson cluster process $\Lambda_0 \subset \mathbb{R}^2$ with intensity $\lambda$ [cluster/m$^2$]. Each cluster $\mathcal{C}_i$ is composed of two terminals. The former, also referred to as cluster centre, is expression of the parent Poisson point process (PPP) and is located at $\mathbf{u}_i$, while the latter is randomly distributed over a circle of radius $r \geq 1$ centred at its companion, i.e., it lies at $\mathbf{v}_i = \mathbf{u}_i + \mathbf{w}_i$, with $\mathbf{w}_i = re^{j\varphi_i}$ and $\varphi_i \sim \mathcal{U}[0, 2\pi)$. Time is divided in slots of equal duration, and every time unit sees a random scattering of the position of all the nodes. More precisely, each cluster centre $\mathbf{u}_i$ is relocated following the high-mobility random walk model presented in [4], [5], whereas the angular component $\varphi_i$ of its peer is independently re-drawn. Under these assumptions, the topology at a generic slot $t$ is described by an independent and homogeneous Poisson cluster process $\Lambda_t$ of intensity $\lambda$ [5].[1]

Within each slot, a node generates a packet for its cluster-fellow with probability $a$. Upon creation, packets are enqueued in a FIFO buffer of infinite size. The transmission of a data unit takes exactly one slot, and the packet is removed from a queue only if correctly received by its destination. In case of a non-empty buffer, a terminal accesses the medium in the current slot with probability $q$, or remains silent with probability $1 - q$, i.e., the network is operated following a $q$-persistent slotted ALOHA policy.[2] Given the uncoordinated nature of the considered MAC, each cluster may experience within a time unit 0) no packet exchange at all, 1) a single transmission from one of its composing terminals, or 2) two simultaneous delivery attempts. In the following, we denote the probabilities of the corresponding events as $p_0$, $p_1$, and $p_2$. Moreover, in case both cluster nodes concurrently send a packet to each other, a FD connection takes place, for which, unless otherwise specified, we assume perfect IC.

Leaning on the homogeneity of $\Lambda_t$, we can characterise the performance of the system focussing on a typical receiver located at the origin of the plane, for which we denote the incoming power from a transmission originated at $\mathbf{u}$ as $P_{\mathsf{rx}}(\mathbf{u})$. All links in the network are affected by path-loss with exponent $\alpha > 2$ and block Rayleigh fading. Accordingly, $P_{\mathsf{rx}}(\mathbf{u}) = PL(\mathbf{u})\zeta$, where $P$ is the transmission power common to all nodes, $L(\mathbf{u}) = u^{-\alpha}$, and $\zeta$ is an exponential r.v. with unit mean, drawn independently for any link and over each time slot. In view of the interference-limited nature of the network under consideration, we disregard thermal noise and abstract the physical layer resorting to a threshold model. More specifically, a packet is received correctly by its addressee as soon as the signal to interference ratio (SIR) is above a reference value $\theta$, which embeds coding and modulation aspects. Recalling the assumption of ideal IC, for any data exchange we can write $\text{SIR} = P r^{-\alpha} \zeta / \mathcal{I}$, where $\mathcal{I}$ accounts for the aggregate interference experienced at the receiver and the ratio does not depend on the employed power $P$.

---
[1]The stability of static HD networks was instead recently tackled in [6].
[2]Within every slot, the displacement of the nodes takes place first. Then, the medium access decision is made prior to packet generation. Thus, each packet has a service time in queue of at least one slot.

*Remark 1:* For the discussed mobility and channel models, the SIR values experienced at a node over subsequent time slots are described by i.i.d. random variables [5].

Within this framework, we focus on characterising the stable behaviour of a FD network, which we introduce as follows.

*Definition*: Let $N_i(t)$ be the number of packets in the buffer of node $i$ when the MAC decision is made for slot $t$. Denoting by $\{N_i(t)\}$ the state of all the queues in the network, we say that the system is *stable* if the Markov chain describing their evolution admits a limiting distribution for $t \to \infty$ [7].

## III. STABILITY REGION OF A FD NETWORK

The stability of buffered Aloha is well-known to be a daunting problem, due to the interdependent evolution of queues within the network. For clustered topologies this aspect is further exacerbated, as the success probability of a sender is determined not only by the aggregate interference at its addressee, but also on whether the latter is concurrently transmitting. In our framework, however, the following holds.

*Remark 2:* When perfect IC is assumed, the SIR at a receiver does not depend on whether it is involved in a half- or full-duplex link. Thus, upon transmission, the probability that a packet be removed from a node's queue is independent of the MAC behaviour of its cluster peer over the same slot.

Leaning on Rem. 1 and 2, the stability region for our system can be characterised following the seminal approach in [4]:

*Theorem 1:* An Aloha-based FD network with ideal IC is stable if and only if the arrival rate at each node satisfies

$$a < q\, e^{-\lambda q \left(2\Omega_1 + q(\Omega_2 - 2\Omega_1)\right)}, \quad (1)$$

where the ancillary functions $\Omega_1(\alpha, r, \theta)$ and $\Omega_2(\alpha, r, \theta)$ are reported in (2) at the bottom of next page, considering $\Gamma(x) = \int_0^\infty x^{t-1} e^{-x} dt$ and $\ell(u, \varphi) = (u^2 + r^2 + 2ru \cos \varphi)^{-\alpha/2}$.

*Proof:* The proof, omitted for brevity, follows the steps of [4, Th. 1]. It suffices to observe that the dominant network considered in [4], i.e., a system where even users with empty queues access the medium with probability $q$, corresponds in our case to a saturated slotted Aloha system where an ideal FD link is triggered if both cluster peers transmit in the same slot. The packet success probability of the latter was derived in [2], [3], obtaining the expression $\exp(-\lambda q(2\Omega_1 + q(\Omega_2 - 2\Omega_1)))$ which leads to (1). ∎

Let us now focus on a stable network in the asymptotic regime. Based once more on the results of Rem. 1 and 2, the theory of thinning for point processes allows to identify two independent homogeneous Poisson cluster processes $\Lambda_1$ and $\Lambda_2$, of intensities $p_1 \lambda$ and $p_2 \lambda$, that track node pairs where a single or two simultaneous transmissions are active over the slot of interest, respectively. By other words, the interference level $\mathcal{I}$ at a typical receiver is equivalent to the one experienced in a system where a fraction $p_1$ of the clusters operates in HD mode, whereas another fraction $p_2$ triggers bi-directional FD links. This parallel allows us to express the probability of successfully decoding a packet through well-established stochastic geometry results [2], [3]:

$$p_s = e^{-\lambda (p_1 \Omega_1 + p_2 \Omega_2)}. \quad (3)$$

In turn, the values of $p_1$ and $p_2$ driving (3) depend both on the persistency parameter $q$ and on the state of the queues within a cluster when the access decision is made. Simple combinatorial arguments bring

$$\begin{aligned} p_1 &= 2q\pi_0(1-\pi_0) + 2q(1-q)(1-\pi_0)^2 \\ p_2 &= q^2\,(1-\pi_0)^2, \end{aligned} \quad (4)$$

where $\pi_0$ is the stationary probability that a terminal has an empty queue at the beginning of a slot. In the former equation, the two addends account for the case of having a single transmission when only one or both nodes have packets to transmit, respectively, whereas $p_2$ stems from the fact that a FD connection requires enqueued data units at the two ends of a cluster. On the other hand, to characterise the behaviour of a queue we observe that, for a stable network, the buffer of a typical node evolves as a Geo/Geo/1 system with arrival rate $a$ and departure probability $qp_s$. A Markovian analysis readily provides both the sought stationary probability $\pi_0 = 1 - a/(q\,p_s)$ and the average number of packets in the queue $N = a(1-a)/(q\,p_s - a)$. Leveraging the expression of $\pi_0$ and plugging (4) into (3) we can eventually express the packet success probability for the stable modelled FD system:

$$p_s = \exp\left(-\lambda\left(2\Omega_1 \frac{a}{p_s} + (\Omega_2 - 2\Omega_1)\frac{a^2}{p_s^2}\right)\right). \quad (5)$$

This is a nonlinear and implicit equation in $p_s$. It is possible to prove that in a stable system three solutions exist, but the two smallest ones are rejected out of physical reasons, since for them higher $a$ (i.e., harsher interference) would lead to higher values of $p_s$. Incidentally, we also note that outside the stability region, the queues are asymptotically empty with zero probability. Thus, classic results for a saturated FD network hold [2], [3], and the success probability after a transient, i.e., for $t \to \infty$, is $\exp(-\lambda q(2\Omega_1 + q(\Omega_2 - 2\Omega_1)))$.

Finally, combining the key result in (5) with the outcome of the Geo/Geo/1 model on the queue size $N$, the average delay $D_{\text{fd}}$ for a stable network follows from Little's law as

$$D_{\text{fd}} = (1-a)/(q\,p_s - a). \quad (6)$$

## IV. A REFERENCE HALF-DUPLEX SCENARIO

The analysis of Sec. III triggers a natural question on the potential of FD in large topologies with respect to a simpler HD counterpart. From this standpoint, we remark that existing HD literature [4] considers a *classical bipolar model* where only one terminal within a cluster generates traffic, whereas the other acts as a receiver. Such a scenario is not the appropriate counterpart for our FD system, since, for a given density $\lambda$, the amount of traffic injected in the network would be intrinsically different in the two cases. To overcome this discrepancy, we remain consistent with the model of Sec. II, yet instantiate the HD constraint enforcing time sharing among peers within a node-pair. More specifically, we still assume that each node generates traffic in every slot with probability $a$. Moreover, the cluster centre can access the medium only at even-indexed slots and acts as a receiver otherwise, while its cluster-fellow is allowed to send only at odd-indexed time units and listens for possible incoming data on even-indexed ones. Within slots where its transmissions are permitted, a node continues to implement the discussed $q$-persistent Aloha.

*Remark 3:* For the considered HD network, the queues of the two users in a cluster are decoupled, i.e., the probability of removing a packet from a node's buffer is independent of the behaviour or state of its peer. Moreover, if we consider the system either at odd- or even-indexed slots only, its behaviour is captured by a *classical bipolar model* through a Poisson clustered point process of intensity $\lambda$.

*Lemma 1:* A HD network with time sharing is stable if and only if the arrival rate at each node satisfies

$$a < \frac{1}{2} q\, e^{-\lambda q\, \Omega_1}. \quad (7)$$

*Proof:* Since the network evolves independently on odd- and even-indexed slots, it suffices to prove the result for either set. For any of them, the system can be described as a family of discrete time queues with time unit equal to two slots. Rem. 3 allows then to apply directly the result of [4, Th. 1], recalling that in our case the average arrival rate per time unit is $2a$ and that the system is noise-free. ∎

Assume now that the network is operated in stable conditions, and consider a typical node. In order to derive the stationary probability $\pi_{0,\text{hd}}$ that its buffer is empty at the time in which a MAC decision is made, we can model the queue as a discrete time G/Geo/1 system that evolves every two slots (from now on referred to as a *HD-unit*). In this case, the probability that a packet under service leaves the queue at the current HD-unit is $q p_{s,\text{hd}}$, where $p_{s,\text{hd}}$ indicates the packet success probability in the stable network. On the other hand, the arrivals within a HD-unit may be either 0, 1 or 2 with probability $(1-a)^2$, $2a(1-a)$ and $a^2$, respectively. Under these hypotheses, discrete-time queueing models [8] allow to compute $\pi_{0,\text{hd}} = 1 - 2a/(qp_{s,\text{hd}})$. On the other hand, recalling Rem. 1 and 3 the success probability of a typical receiver is the one of a *classical bipolar model* with intensity $\lambda q(1-\pi_{0,\text{hd}})$, derived in [5]. Combining these results we have

*Corollary 1:* The stable success probability for our HD network is $p_{s,\text{hd}} = \exp(-2\lambda\Omega_1 \cdot a/p_{s,\text{hd}})$.

Finally, notice that the G/Geo/1 model considered so far does not suffice to capture the queueing delay $D_{\text{hd}}$, as the latter intrinsically depends on the specific slot within the HD-unit in which the packet was created. Moreover, the HD constraint induces a statistic of the service time for newly generated data that is dependent on whether it encounters an empty or non-empty buffer. We instead provide an upper bound for $D_{\text{hd}}$, whose tightness will be verified in Sec. V.

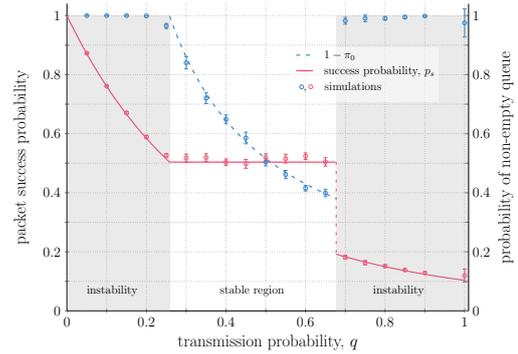

Fig. 1. Probability of non empty queue and $p_s$ vs $q$. $\lambda = 0.2$, $a = 0.13$. Lines represent analytical solutions, while points validating simulations.

*Corollary 2:* The average delay $D_{\text{hd}}$ in slots for a stable HD network satisfies

$$D_{\text{hd}} \leq (3a-2)/(2a - qp_{s,\text{hd}}). \quad (8)$$

*Proof:* The upper bound is obtained assuming that a packet generated in a HD-unit can start service only during the following one. It is easy to verify that under this hypothesis the service times are multiples of two slots following an i.i.d. geometric distribution, i.e., $\Pr\{\text{service time of } 2k \text{ slots}\} = qp_{s,\text{hd}} \cdot (qp_{s,\text{hd}})^{k-1}$, $k \in \mathbb{N}^+$. For this G/Geo/1 system the delay can be derived working with probability generating functions [8], obtaining the bound in (8). ∎

## V. RESULTS DISCUSSION

The analytical framework was verified through Matlab simulations. Unless otherwise stated, $\alpha = 4$, $\theta = 2$, $r = 1$, and each point averages 40 runs. Moreover, the statistical significance is displayed through 95% confidence bars.

After a transient, Fig. 1 depicts for a FD network the probability that the queue of an average node is non empty as well as the success probability $p_s$, thus validating the theoretical results from Sec. III. To interpret the trends, it is useful to recall the stability condition offered by (1). For a given $(\lambda, a)$ pair, in fact, the network traverses three phases when the access probability increases. For low values of $q$, the system is unstable despite a high success probability simply because nodes are too conservative in emptying their queues. At the other end, a too aggressive access increases excessively the level of interference, causing instability due to repeated delivery failures. Within the stable region, finally, a change in the persistence parameter does not alter $p_s$, thus suggesting to employ the highest possible $q$ so to reduce the service delay.

A first non-trivial question is whether the stability region can be extended by means of FD, recalling the tradeoff between spatial reuse and additional interference. To answer this, Fig. 2 reports the maximum stable arrival rate $a^*$ for a HD- and a

$$\Omega_1 = (\pi r^2)\,\theta^{2/\alpha}\Gamma(1-2/\alpha)\Gamma(1+2/\alpha), \qquad \Omega_2 = \int_0^\infty 2u\left(\pi - \frac{1}{1+\theta r^\alpha u^{-\alpha}} \cdot \int_0^\pi \frac{1}{1+\theta r^\alpha \ell(u,\varphi)}d\varphi\right)du \quad (2)$$

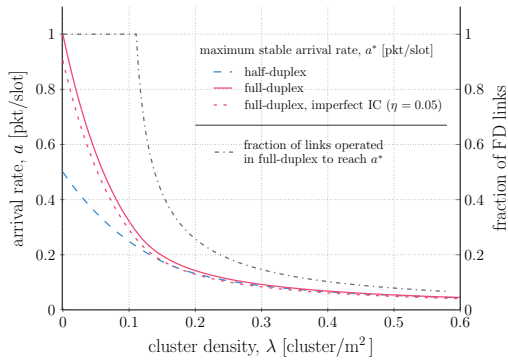

Fig. 2. Maximum stable arrival rate for half- and full-duplex networks against $\lambda$. Also depicted the probability that an active cluster instantiates a FD link.

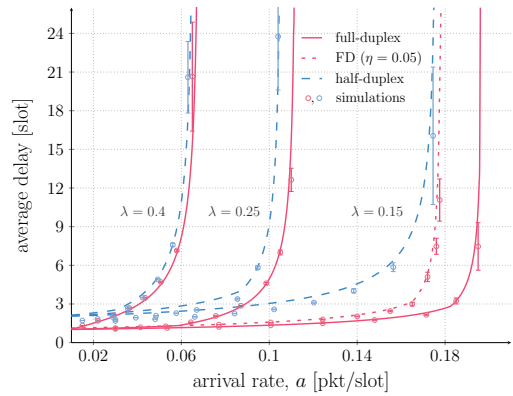

Fig. 3. Average delay vs $a$. For HD, the analytical curve reports the bound in (8), whereas simulation exactly implement the system described in Sec. IV.

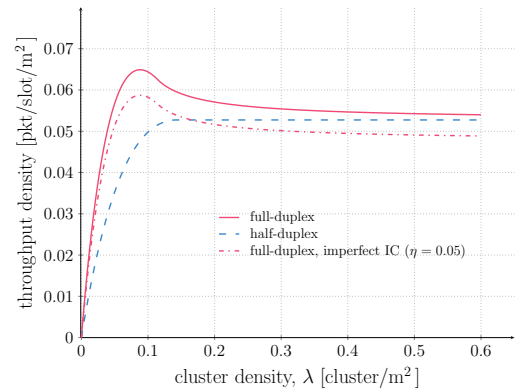

Fig. 4. Maximum throughput density vs $\lambda$. The dash-dotted line reports the behaviour of FD with imperfect IC and $\eta = 0.05$.

FD-system against $\lambda$, obtained analytically from (7) and (1).[3] Remarkably, the plot prompts FD as particularly beneficial for sparse networks. Conversely, for larger $\lambda$ the maximum stable arrival rate converges to the one of a HD system. To understand why, we report the probability that an active cluster instantiates a bi-directional connection, i.e., $p_2/(p_1 + p_2)$. In dense networks, the interference level is structurally high, and FD exchanges would undermine the SIR. Thus, the optimal access probability rapidly adapts, inducing very few FD links. In such conditions, a HD system may be the best option. We further elaborate on this considering the role played by imperfect IC. Leaning on a well-established approach, we model residual self-interference assuming that a fraction $\eta \in (0, 1)$ of the transmission power adds up to the denominator of the SIR when a FD connection is established. Accordingly, the success probability of a saturated network is simply scaled by a factor $\beta = \exp(-\eta \theta r^\alpha)$ [2]. In turn, the assumption would again correlate nodes' queues, since the success probability at a terminal changes if its peer is concurrently sending data. To overcome this, we focus on a lower bound for the FD system, where the SIR at a receiver is always affected by an additional noise $\eta P$, regardless of whether the same node is also transmitting. In such a setting the proof of Th. 1 can be repeated, obtaining a stability region that is simply scaled by a factor $\beta$ with respect to ideal IC, as shown in Fig. 2.

A second relevant metric is the delay in (6) (8), reported by Fig. 3 against $a$ for various $\lambda$. The curves diverge at the respective $a^*(\lambda)$, and the gap between the HD and FD asymptotes grows for lower $\lambda$. The delay advantage is remarkable for $\lambda = 0.15$, where the ideal FD network performance is consistently at least twice as good the HD system, shoring up again the benefits in sparse topologies. On the other hand, the plot also shows for $\lambda = 0.15$ how imperfect IC may in fact disrupt any achievable gain.

Finally, we study the FD throughput edge over HD, one of the major selling points of this technology. Fig. 4 compares the maximum throughput density in stability conditions, i.e., $\tau^*(\lambda) = 2\lambda a^*(\lambda)$. The HD system performance saturates since $a^*$ decreases $\propto 1/\lambda$ for large $\lambda$. The FD system is able for sparse networks to double the throughput density, but the head start tapers down since, as discussed, less aggressive access probabilities make the FD network behave more like a HD one. Let us remark that the throughput achieves a maximum for $\lambda \simeq 0.09$ (as opposed to the asymptotic maximum in the HD case). Fig. 2 reveals that at such cluster density interference becomes a limiting factor even in cases where it is optimal for all links to be operated in FD mode. Under imperfect IC, the scaled version of the capacity region discussed earlier leads to a lower bound on the throughput density as $2\beta\lambda a^*(\lambda)$, reported in Fig. 4. The plot confirms the trends presented so far, highlighting the critical role of IC in dense networks.

## REFERENCES

[1] A. Sabharwal, P. Schniter, D. Guo, D. Bliss, S. Rangarajan, and R. Wichman, "In-Band Full-Duplex Wireless: Challenges and Opportunities," *IEEE Journ. Sel. Areas Comm.*, vol. 32, no. 9, pp. 1637–1652, Sep. 2014.
[2] Z. Tong and M. Haenggi, "Throughput Analysis for Full-Duplex Wireless Networks with Imperfect Self-interference Cancellation," *IEEE Trans. Comm.*, vol. 63, pp. 4490–4500, Nov. 2015.
[3] A. Munari, P. Mähönen, and M. Petrova, "A Stochastic Geometry Framework for Asynchronous Full-Duplex Networks," 2015. [Online]. Available: http://arxiv.org/abs/1512.01478
[4] K. Stamatiou and M. Haenggi, "Random-Access Poisson Networks: Stability and Delay," *IEEE Comm. Lett.*, vol. 14, no. 11, Nov. 2010.
[5] F. Baccelli and B. Błaszczyszyn, *Stochastic Geometry and Wireless Networks, Volume I - Theory*. NoW Publisher, 2009.
[6] Y. Zhong, M. Haenggi, T. Quek, and W. Zhang, "On the Stability of Static Poisson Networks under Random Access," *IEEE Trans. Comm.*, vol. 64, Jul. 2016.
[7] W. Szpankowski, "Stability Conditions for Some Distributed Systems: Buffered Random Access Systems," *Adv. Appl. Prob.*, vol. 26, 1994.
[8] H. Bruneel, "Performance of Discrete-Time Queueing Systems," *Computers and Operations Research*, vol. 20, no. 3, pp. 303–320, Apr 1993.

[3]Strictly speaking, the points do not belong to the respective stable regions. With a slight abuse of notation, the results are intended for $a = a^* - \varepsilon$, $\varepsilon > 0$.